# A HYBRID CLASSIFICATION ALGORITHM TO CLASSIFY ENGINEERING STUDENTS' PROBLEMS AND PERKS

Mitali Desai and Mayuri A. Mehta

Department of Computer Engineering,
Sarvajanik College of Engineering and Technology, Surat, India

## ABSTRACT

*The social networking sites have brought a new horizon for expressing views and opinions of individuals. Moreover, they provide medium to students to share their sentiments including struggles and joy during the learning process. Such informal information has a great venue for decision making. The large and growing scale of information needs automatic classification techniques. Sentiment analysis is one of the automated techniques to classify large data. The existing predictive sentiment analysis techniques are highly used to classify reviews on E-commerce sites to provide business intelligence. However, they are not much useful to draw decisions in education system since they classify the sentiments into merely three pre-set categories: positive, negative and neutral. Moreover, classifying the students' sentiments into positive or negative category does not provide deeper insight into their problems and perks. In this paper, we propose a novel Hybrid Classification Algorithm to classify engineering students' sentiments. Unlike traditional predictive sentiment analysis techniques, the proposed algorithm makes sentiment analysis process descriptive. Moreover, it classifies engineering students' perks in addition to problems into several categories to help future students and education system in decision making.*

## 1. INTRODUCTION

Social media has emerged as a highly useful personal communication technology which provides a large platform for individuals to share their feeling, emotions, and opinions in an informal and casual way to seek social support. The opinion data available on internet covers all the recent trends, issues, opinions on every thinkable topic [1-2]. The students' informal talks on social media platforms such as Twitter, Facebook, and YouTube shed insight into their present or past learning experiences. Mining such homogeneous data extracts useful pattern from the large volume of students' generated raw data to support decision making in education system.

The education data mining involves two broad categories of research: front stage study and back stage study [3]. Front stage study is usually conducted in the controlled class room environment by making student fill the questionnaires or feedback forms. The defined format of answers, the guarded ambience and requirement of spontaneous answers are the shortcomings of this method [1-4]. Moreover, the collected responses have structured format that makes the outcome of the study predictable. Additionally, front stage study techniques are very time consuming, especially when a number of students is large and hence, the analysis process cannot be carried out frequently [3-4]. Moreover, they provide limited scalability [5]. On the other hand, the back stage study provides relaxed environment allowing students to share their honest opinions on the social media platform. Such data provides concrete and essential knowledge about students' emotions toward the learning experiences.

Social media serves as a substantial platform to provide large input data for research. Therefore, the majority of the analytics use social media platforms for data collection. However, the abundance of data, variety of platforms, use of slangs, timing of different posts on web, and diversity of language make the analysis process difficult to gain the hidden emotions in educational decision making [3-4]. Hence, an efficient analysis and classification technique is required that incorporates qualitative analysis and automated technique to get deeper insight into the students' sentiments.

The existing classification techniques are useful in disaster relief and humanitarian assistance, marketing predictions, checking customer loyalty, finding job opportunities, population health care and understanding students' learning experiences [1-7]. The recommendation systems, websites that disclose job opportunities, push notification and Google health care applications are some of the examples. The educational system is still lagging behind in the field of descriptive sentiment analysis. To the best of our knowledge, very few attempts have been made to mine and classify students' data generated on uncontrolled social media space to get deeper understanding of their problems along with their perks.

In this paper, we propose a novel Hybrid Classification Algorithm (HCA) for descriptive sentiment analysis to understand students' problems and perks deeply. We integrate both subjective analysis and data mining techniques to make the process descriptive. The proposed algorithm classifies engineering students' sentiments into several categories that help future students and education system in decision making. We have developed a dynamic method to generate various categories. The dynamic process eliminates the requirement of changing the algorithm for newly added data. In our study, we have collected data from Twitter that has witnessed a tremendous growth in the number of users recently [8-9]. The Tweets of engineering students on Twitter consist of #engineeringProblems and #engineeringPerks hashtags. Tweets are mostly public and limited to 140 characters that simplify identification of emotions in text [9-11].

The rest of the paper is organized as follows: In section 2, we discuss the existing work related to sentiment analysis. Section 3 describes the proposed approach. Finally, Section 4 specifies the conclusion and future directions.

## 2. RELATED WORK

In recent years, a voluminous amount of research has been done in the field of sentiment analysis. In [12], authors have presented the logical approach for extraction of the sentiment on widely used social networking sites. They have analysed the sentiments of the text using combinatory categorical grammar, lexicon acquisition and annotation, and semantic networks for analysing. The basic techniques of sentiment classification and the methods for data collection are presented in [13]. Classification accuracy of the feature vector is tested for electronic products domain using different classifiers such as Nave Bayes, SVM, Maximum Entropy, and Ensemble classifiers in [14]. In [15], authors have introduced a hybrid method that combines usage of sentiment lexicons along with a machine learning classifier for polarity detection of opinionated texts in the domain of consumer-products. In [16], authors have proposed a set of techniques of machine learning with semantic analysis for classifying the sentence and product reviews based on twitter data using WordNet for better accuracy. In [17], authors have examined the performance of different classifiers such as Naive Bayesian, SMO, SVM and Random Forest to classify Twitter data.

From our in-depth study on the existing techniques, it is observed that the existing techniques typically classify the text data into pre-set categories: positive, negative and neutral. Classifying the students' sentiments into positive or negative category does not provide deeper insight to their problems. In [7], authors have proposed a technique to classify students' data generated on Twitter into various categories. However, the category generation process is static. Moreover, the static process provides limited accuracy of the classifier. Our proposed algorithm provides the dynamic method to generate various categories. The dynamic process eliminates the requirement of changing the algorithm for newly added data. Moreover, it increases the accuracy of the classifier. Additionally, the proposed algorithm identifies students' perks in addition to problems.

## 3. THE PROPOSE HYBRID CLASSIFICATION ALGORITHM

The sentiment analysis of student data is an emerging field that needs much more attention. In addition, classifying students' emotions into just positive or negative opinions do not shed any light into the actual problems in the existing learning process. The proposed hybrid classification algorithm classifies students' problems and perks shared on Twitter into various belonging categories rather than merely positive or negative. Thus, the proposed algorithm makes the sentiment analysis process descriptive. The descriptive technique is more useful to get deeper insight into students' problems and perks than the traditional predictive techniques. Figure 1 shows the processing steps of hybrid classification algorithm.

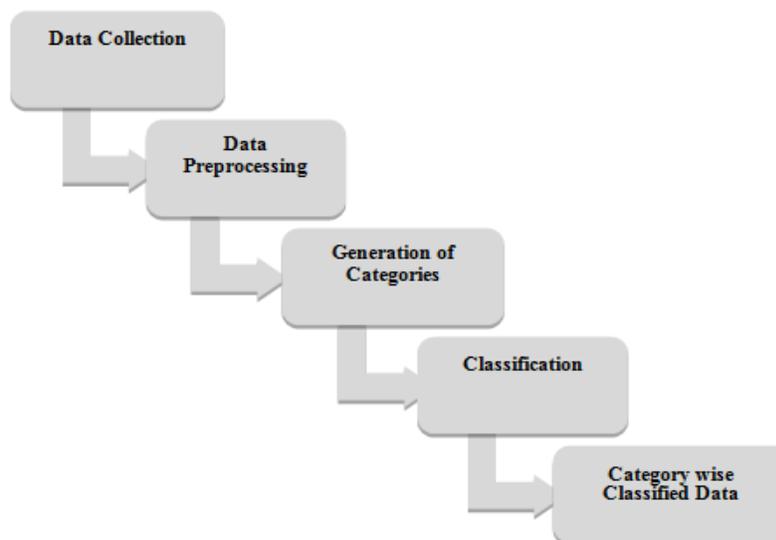

Figure 1. Processing steps of hybrid classification algorithm

### 3.1 Data Source

Selection of data source to conduct the back stage study to classify students' problems and perks plays significant role. Social media platform as the data source are categorize into three general categories: blogs, micro-blogging sites, and review site [12-15]. Among all categories, a micro-blogging site such as Twitter has gained higher popularity due to its limited strength of the content. Hence, in our study, we have collected students' data from Twitter.

The message posted on Twitter is called Tweet. Tweets are limited to 140 characters. They are composed of any of the followings [16] [18-19]: text, links, images, and six seconds video. The mining process is applied to classify these components into positive or negative categories. The Tweets contain three notations which are hashtags (#), retweets (RT), and account Id (@).

There are three possible ways to collect Twitter data for research as follows.

- Data repositories such as UCI, Friendster, Kdnuggets, and SNAP

- Application Program Interfaces (APIs): Twitter provides two types of APIs such as search API and stream API

- Automated tools that are further classified into premium tools such as Radian6 [18], Sysmos, Simplify360, Lithium, SproutSocial and non-premium tools such as Keyhole, Topsy, Tagboard and SocialMention

Amongst above mentioned tools, Topsy is a Graphical User Interface (GUI) based non premium tool that makes data extraction process non-tedious. Hence, we have collected Twitter data of engineering students consisting of #engineeringProblems and #engineeringPerks hashtags using automated tool Topsy.

### 3.2 Data Pre-processing

The collected data is raw data. In order to apply classifier it is necessary to pre-process the raw data. The pre-processing task involves uniform casing, removal of hashtags and other Twitter notations (@, RT), stop words, and decompression of slang words.

### 3.3 Generation of Categories

From the qualitative analysis of the collected data, we generate different categories to classify students' problems and perks. In order to make the category generation process dynamic, we construct the corpus. The corpus consists of all the possible words from the dataset which encompasses a single class category.

### 3.4. Sentiment Classification Techniques

There are typically two techniques to identify sentiment of the text [7] [12-13] [18-19]: knowledge based method and machine learning methods. To make our classification algorithm hybrid, we combine both classification techniques. Hybrid approach increases the accuracy of classifier and makes the category generation process dynamic. Firstly, we create a corpus to make the category generation process dynamic using knowledge based method. The dynamic category generation process eliminates the requirement of changing the algorithm for newly added data. Secondly, we apply machine learning method to classify the input data into several problems and perks categories.

Knowledge based method is also called Lexicon based method. The lexicon-based approach depends on finding the opinion lexicons that are used to analyse the text. This method requires preparation of dictionary or corpus containing all the possible opinion words with their polarities i.e. positive or negative. It finds opinion seed words and subsequently, it searches the dictionary of their synonyms and antonyms. A small set of opinion words is collected manually with known orientations. The set is grown by searching in the well-known lexicon dictionary tool such as WordNet or Sentiful for their synonyms and antonyms [16-17].

Machine Learning methods make machine to learn from the past experiences and to behave like a human. They are further classified into supervised and unsupervised methods [12] [14-24]. To carry out sentiment analysis, typically the supervised machine learning methods are used. Supervised learning methods highly depend on training set and test set. Training set is already labelled data whereas test set is used to validate performance of the built classifier. A large number of supervised machine learning algorithms such as Naïve Bayes, Logistic Regression, Support Vector Machine, Maximum Entropy, Decision Tree, Random Forest, and Bayesian Network are used for sentiment analysis [7] [12-22]. Choice of an appropriate algorithm for selected domain is a crucial step. It has been observed that the following machine learning algorithms are widely used and give average accuracy in majority of domains as well as with different types of data. Moreover, they provide consistent speed of classification process irrespective of the size of input data while handling the outliers

1) Naïve Bayes (NB) Approach

Naïve-Bayes classifier [7] [17-20] is a simple probabilistic classifier which uses Bayes Theorem and finds maximum likelihood probability of a word belonging to a particular class. The probability P is defined as follows:

$$P\left(Xi \mid c\right) = \frac{Count\ of\ Xi\ in\ document\ of\ class\ c}{Total\ no\ of\ words\ in\ document\ of\ class\ c} \qquad (1)$$

Naïve -Bayes classifier suffers from an assumption that all the features in the feature space are independent. The frequency counts of the words are stored in hash tables during the training phase. As per the definition of probability, the document d is classified into class c using following equation:

$$c \ast= argmax\ P\left(c \mid d\right) \qquad (2)$$

2) Maximum Entropy

The maximum entropy technique is a probability distribution estimation technique. It is used for various natural language processing tasks such as text classification. It depends on the probabilistic approach like Naïve Bayes [21-22]. The underlying principle of maximum entropy is that if more information regarding the data is not known, the distribution should be as uniform as possible. This constraint allows the distribution to be minimally non uniform. The probability is derived from the labelled training data and is represented as expected values of features as follows:

$$P\,(c\mid d) = \frac{1}{Z\,(d)\{\exp\,(\sum \lambda i\; fi\,(c,d))\}} \tag{3}$$

Where fi(c,d) is a feature, $\lambda i$ is a parameter to be estimated and Z(d) is a normalization function. Unlike NB, maximum entropy doesn't make any assumption regarding the feature dependency. The motivating idea behind maximum entropy is that one should prefer the most uniform models that satisfy any given constraints.

3) Support Vector Machine (SVM)

The main principle of SVMs is to determine a linear separator that separates the different classes in the search space with maximum distance [12-16]. If we symbolize the hyper plane as h, the tweet as t, and represents the classes as Cj € {l, -1} into which the tweet has to be classified, the solution can be written as follows corresponding to the sentiment of the tweet.

$$\vec{h} = \sum i\; ai\; Ci\; \vec{tj}\,, \qquad ai \geq 0 \tag{4}$$

In our Hybrid Classification Algorithm (HCA), we apply SVM to classify the input data into several categories since SVM provides higher theoretical accuracy in terms of correctly classifying the input data into valid categories. Moreover, it provides better speed of classification process compare to other algorithms. Additionally, SVM supports the maximum margin technique by separating two data points with the maximum possible distance to avoid overlapping. Figure 2 shows the hybrid classification algorithm.

---

**Input:** a dataset T = {$t_1$, $t_2$, $t_3$, ……., $t_n$} of n Tweets that consist of #engineeringProblems and #engineeringPerks hashtags
**Output:** classification of n Tweets into m categories $c_1$, $c_2$, $c_3$, …., $c_m$

Hybrid Classification Algorithm (T, n)

**BEGIN**
1.  for i = 1 to n Tweets
2.      Convert $t_i$ into uniform case
3.      Remove Twitter notations, emoticons, URLs, and stop words from $t_i$
4.      Compress the elongated words in $t_i$
5.      Decompress the slang word in $t_i$
6.  Generate $c_1$, $c_2$, …, $c_m$ dynamically into corpus using lexicon method
7.  Apply SVM to classify n Tweets into their respective belonging class categories
**END**

---

Figure 2. The proposed hybrid classification algorithm

In HCA, first n Tweets of Engineering students consisting of #engineeringProblems and #engineeringPerks hashtags are collected. Subsequently, the collected data is pre-processed applying steps 2-5. In pre-processing, first the Tweets are converted in uniform casing i.e. either upper case or lower case. Then the Twitter notations such as hashtags (#), retweets (RT), and account Id (@) are removed. Moreover, the URLs and emoticon are also removed. It is necessary to remove non letter data and symbols as we are dealing with only text data. In addition, stop words (i.e. are, is, am) are removed. The stop words do not emphasize on any emotions, it is intended to remove them to compress the dataset. Secondly, compress the elongated words such as happyyy into happy. Thirdly, the slag words such as g8, f9 are decompressed. Generally slang words are adjectives or nouns and they contain the extreme level of sentiments. So it is necessary to decompress them. In step 6, we generate the corpus including all possible word belonging to different identified categories. We make the category generation process dynamic by constructing the corpus using lexicon method. In step 7, we apply the supervised machine learning classifier called SVM to classify the n Tweets into their belonging categories. SVM provides higher theoretical accuracy in terms of correctly classifying the input data point into a valid category. Moreover, it provides better speed of classification process compare to other algorithms.

The generated categories shed insight into students' actual existing problems and perks deeply than merely identifying positive or negative emotions. Moreover, they guide future students in determining their educational field. Furthermore, they are useful in improving the quality of education system.

# 4. CONCLUSION AND FUTURE WORK

In this paper, we have proposed a novel Hybrid Classification Algorithm that makes the traditional predictive sentiment analysis process descriptive. To make sentiment analysis process descriptive, we have incorporated qualitative analysis along with data mining techniques. The descriptive process classifies engineering students' Twitter data into their belonging problems and perks categories rather than just positive or negative content identification. The descriptive categories guide the future students in their selection of educational stream. Moreover, based on the descriptive categories, the quality of education system can be improved further. In addition, the proposed algorithm makes the category generation process dynamic that eliminates the need to change the algorithm each time when new data is added. The dynamic category generation processes increase the accuracy of the classifier. In future, we aim to evaluate the performance of the HCA under different data collection methods and data set of different sizes.

## AUTHORS

**Mitali Desai** received the B.E. degree in Information Technology from Sarvajanik College of Engineering and Technology, Gujarat Technological University, Surat, India in 2012. She is pursuing M.E. in Computer Engineering at Sarvajanik College of Engineering and Technology, Gujarat Technological University, Surat, India. She has worked as a Teaching Assistant in the Sardar Vallabhbhai National Institute of Technology (SVNIT), Surat, India for 2 years from 2012 to 2014. She has also worked as an Ad-hoc lecture in Sarvajanik College of Engineering and Technology, Surat, India. Her research areas are Data Mining, Web Mining, Soft Computing, BigData Analytics and Artificial Intelligence.

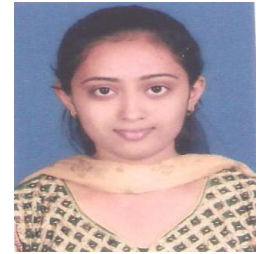

**Mayuri A. Mehta** received the B.E. and the M.E. degrees in Computer Engineering from Sardar Patel University, Vallabh Vidhyanagar, India, in 2000 and 2005, respectively and the Ph.D. degree in Computer Engineering from the Sardar Vallabhbhai National Institute of Technology (SVNIT), Surat, India, in 2014. She started her career as Lecturer in 2000 and presently she is working as Associate Professor with Department of Computer Engineering, Sarvajanik College of Engineering and Technology, Surat, India. Her current research interests include Distributed Computing Paradigms, Analysis and Design of Algorithms and BigData Analytics.

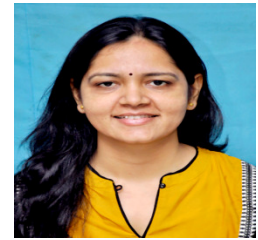